# Improving Positron Lifetime Spectra Quality by Suppressing Corrupted Coincidences via Pulse-Height Spectrum Window Adjustments

**D. Boras**[a,1], **D. Petschke**[a] **and T. Staab**[a]

[a] *University Würzburg, Department of Chemistry, LCTM,
 Röntgenring 11, Würzburg D-97070, Germany*
 *E-mail*: dominik.boras@uni-wuerzburg.de

ABSTRACT: Positron Annihilation Lifetime Spectroscopy (PALS) is a powerful technique for detecting microstructural defects in various material classes. In a commonly used 180° detector configuration equipped with plastic scintillators, measurement accuracy is mainly affected by simultaneous detection of 1275 keV and 511 keV gamma quanta in the same detector. This study introduces an optimization approach that significantly improves spectra quality without requiring hardware modifications. By systematically adjusting the upper threshold of the start window in the Pulse Height Spectrum (PHS), this method effectively reduces unwanted 1275/511 piled-up events. Experimental validation with high purity aluminium samples demonstrates that lowering this threshold effectively suppresses corrupted coincidences, yielding extracted characteristic lifetimes closer to those from a 90° configuration. While a reduced upper threshold slightly broadens the instrument response function (IRF) due to an effectively reduced number of scintillation photons, the overall impact on timing resolution remains negligible. This approach enhances the reliability of PALS measurements while still preserving the benefits of a 180° configuration, i.e. high solid-angle efficiency.

## Contents

## 1. Introduction

Positron Annihilation Lifetime Spectroscopy (PALS) is a highly sensitive and non-destructive technique for analysing microstructures and lattice defects in various material classes, including metals and their alloys [1-5], semiconductors [6-8], and polymers [9, 10]. By utilizing positrons as probe particles, PALS enables the detection and characterization of nano- to mesoscopic open-volume defects such as vacancies [1, 4, 11], dislocations [2, 12], and grain boundaries [13]. These lattice defects are characterized by a locally reduced electron density, which increases the mean positron lifetime thus allowing their identification through lifetime spectrum analysis.

In laboratory-based PALS, $^{22}$Na is commonly used as a positron source. The technique relies on measuring the time difference between the 1275 keV (START) gamma quantum emitted during positron formation from the $\beta^+$ decay and one of the two collinearly emitted 511 keV (STOP) annihilation gamma quanta. The timing of these gamma quanta is measured using photomultiplier tubes (PMTs) with attached scintillators. Scintillator materials widely employed in this setup include inorganic scintillators such as BaF$_2$ [14, 15] and plastic scintillators [16, 17].

Plastic scintillators offer distinct advantages due to their single step scintillation process, rapid rise and decay times, and low effective atomic number which reduces the probability of gamma quantum backscattering[17-19]. Unlike inorganic scintillators, which require alternative source positioning eventually with additional lead shielding to minimize backscattering [20, 21], plastic scintillators are assumed to allow measurements in a 180° face-to-face configuration while covering a larger solid angle which as a consequence reduces the measurement time considerably.

This fundamental assumption of a reduced gamma quantum backscattering rate in plastic scintillators was confirmed in our recent simulation study [18]. We could demonstrate that the occurrences[1] of backscattered 1275 keV quanta (BS-S) is around 1% for a truncated cone geometry of commonly used scintillator dimension and therefore negligible [18].

More importantly, the study revealed that the occurrences[1] of double-detected STOP events (DD), where both collinearly emitted 511 keV quanta strike and directly interact with the opposing scintillators, is by no means insignificant. For the scintillator dimension used in this study (l: 27.9 mm $r_1$: 40.0 mm $r_2$: 19.0 mm), the occurrence[1] rate is approximately 20%, evidently increasing the probability of overlap with the associated 1275 keV quantum (see Figure 1). Other event combinations that could also result in an overlap of at least two events in the START detector can be neglected due to their extremely low probability (Figure 1). Consequently, the dominant contribution to signal overlap in the START detector originates from DD events.

Since the dominant contribution of lifetimes in the spectrum is centred around the t0 channel and thus fall within the broadening range of the Instrument Response Function (IRF) (FWHM: ~200 ps), the majority of these overlapping events are difficult or even impossible to distinguish. This is because their relative time differences, which corresponds to the lifetimes, are significantly shorter than the rise time of plastic scintillator based PMT-pulses, typically ranging from 3 to 5 ns [16, 22]. As this overlap cannot be effectively eliminated through pulse or shape filtering, this event type predominantly contributes to the degradation of the spectrum quality [23]. We hereafter refer to it as a 1275/511 pile-up event.

In this study, we investigate the contribution of these 1275/511 pile-up events to the Pulse Height Spectrum (PHS) through both simulation and experiment. We demonstrate that adjusting the upper limit of the PHS start window directly influences the characteristic lifetimes and the broadening (FWHM) of the IRF extracted from the spectrum. By lowering the upper limit, a significant fraction of this event type can be effectively suppressed, substantially improving the quality of the lifetime spectrum without requiring hardware modifications.

As a reference for ground-truth-like spectrum quality, we use a lifetime spectrum obtained from a 90° configuration, where these event types are fully eliminated. The experiments were conducted using high-purity aluminum (5N5) as the sample material.

---

[1] The term occurrences describe the number of individual interactions with the scintillator of either 1275 keV, 511 keV or combinations of both quantum energies with respect to the total number of recorded scintillator interactions. Therefore, this rate is not related to coinciding interactions in opposing scintillators, i.e. a coincidence of 1275 keV and 511 keV quanta, and consequently does not directly relate to the fraction of corrupted lifetimes contributing to the spectrum. For more information we refer to our recent study [18].

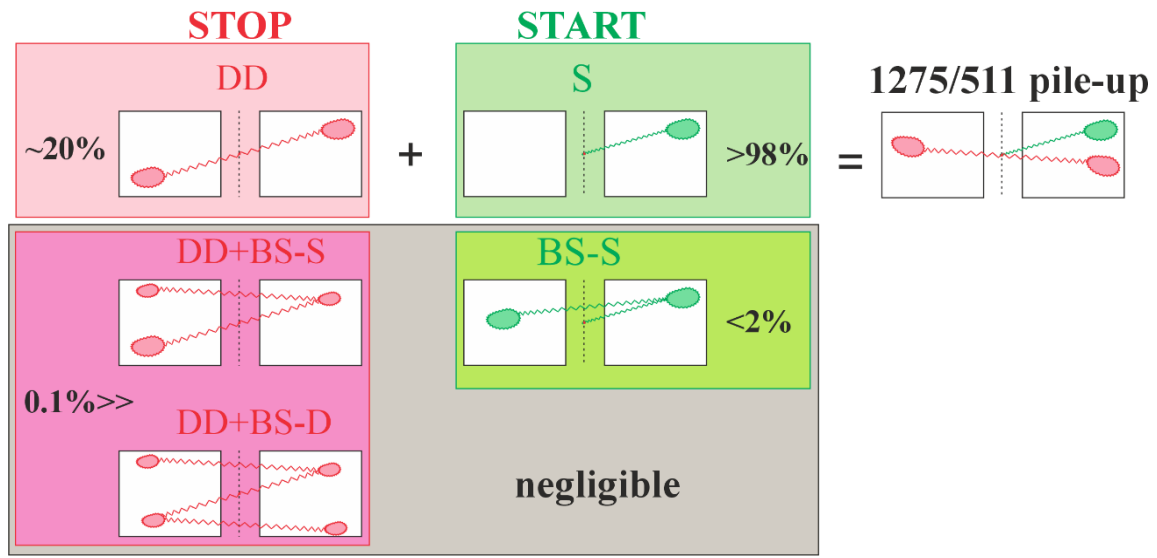

Figure 1: Illustration of the event matrix leading to 1275/511 pile-up occurrences. For the detection of 511 keV gamma quanta, double detections (DD) may occur, accounting for approximately 20% of all 511 keV detections. Additionally, DD events may be accompanied by single backscattering (BS-S) of a 511 keV quantum, or by backscattering of both quanta (BS-D). For the 1275 keV gamma quantum, either a single detection (S) or a single backscattering (BS-S) can occur. However, event combinations such as 511 keV DD+BS-S, DD+BS-D, and 1275 keV BS-S are highly unlikely and can be considered negligible. Therefore, the dominant contribution to 1275/511 pile-up events arises from the combination of 511 keV DD and S of the 1275 keV gamma quantum.

## 2. Degradation of spectrum quality due to 1275/511 pile-up events

As described in the previous chapter, a 1275/511 pile-up event primarily results from the detection of all three gamma quanta associated with positron creation via $\beta^+$ decay (1275 keV) and its subsequent annihilation with an electron (511 keV). In this process, both collinearly emitted 511 keV quanta are simultaneously detected in the opposing detectors, while one of these detectors also registers the associated 1275 keV quantum. Consequently, the detector responsible for the START signal experiences pulse overlap of the 1275 keV and a 511 keV gamma quantum. This overlap not only results in an increased area and amplitude, extending the upper region of the PHS to higher values, but even more importantly adversely affects the accurate determination of the timing signal when using the Constant Fraction (CF) principle as illustrated in Figure 2.
The extent of signal overlap depends on the lifetime, i.e. the time difference between the two gamma quanta being registered in the same detector. If the time difference of the 1275/511 pile-up events is within a few sampling points of the digitizer (approx. a few ten picoseconds), the signals sum to form an apparently single and undistorted pulse (see Figure 2A). Those kinds of pulses cannot be eliminated using standard pulse filters. However, if the time difference is greater, approximately within the rise time or longer, the summed signals exhibit visible deformation in the rising edge as illustrated in Figure 2B. Such deformations can be clearly identified using pulse filters, such as area or pulse-form filters and can therefore be suppressed [24-26].

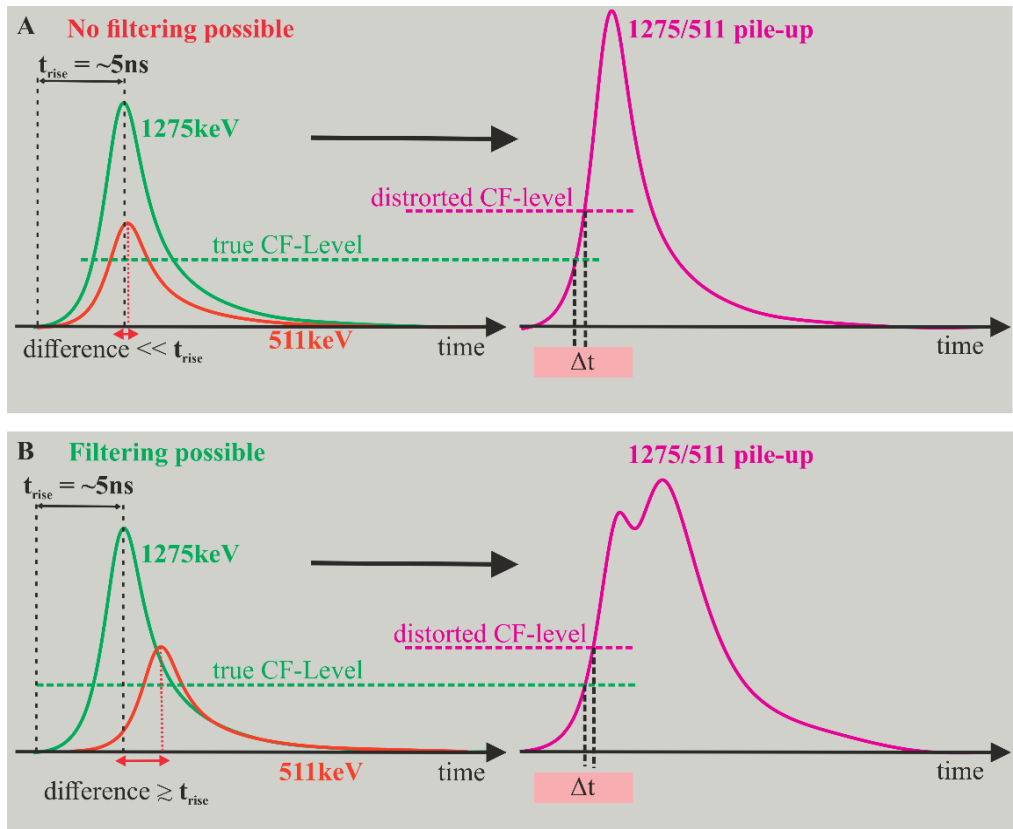

Figure 2: Illustration of the generation of overlapping signals caused by 1275/511 pile-up events detected in a single detector. (A) shows a 1275 keV gamma quantum and a 511 keV annihilation gamma quantum, emitted almost simultaneously with a time difference corresponding to a short positron lifetime in the range of a few sampling points (approx. a few 10 ps). This scenario leads to a distortion of the timing signal (Δt) when using the CF principle. Moreover, these events cannot be supressed by filtering techniques. (B) demonstrates the generation of such events were the time difference between the summed 1275keV and 511keV gamma quanta is in the range of the rise time ($t_{rise}$). The summation of the signals causes a visual deformation and, therefore, these kinds of pile-up events are possible to be supressed by pulse-shape or area filtering techniques [24-26].

Given that the majority of positrons lifetimes forming the spectrum relate to time differences typically below 1 ns (e.g. for samples such as metals and their alloys or semiconductors), the contribution of 1275/511 pile-up events which can be potentially eliminated through filtering techniques is only a few percent [23].

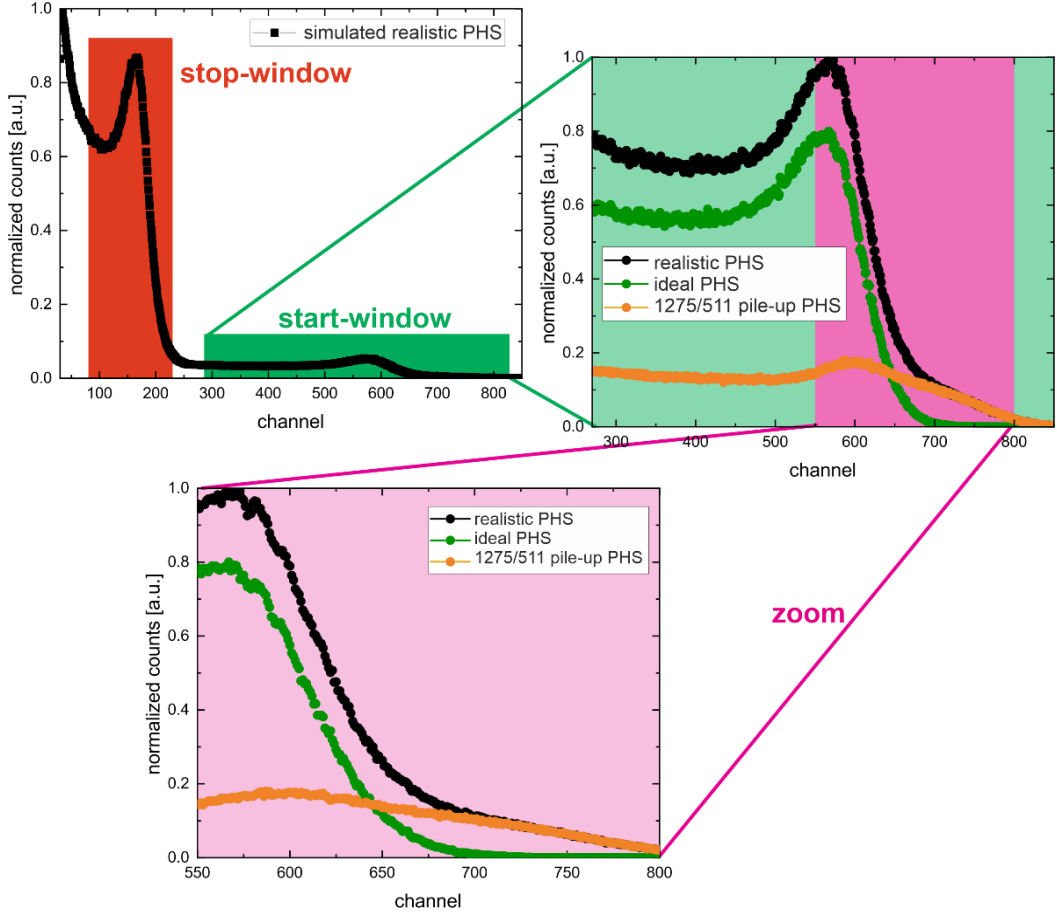

Figure 3: Representation of a simulated PHS using Geant4 based on parameters used in the experimental part of this study. The pulseheight distribution for 1275/511 pile-up events (orange dots) are displayed alongside those of an ideal PHS (green dots), where exclusively 1275 keV gamma quanta were considered. The resulting PHS, representing the realistic PHS is depicted by black dots and combines both PHS from the 1275/511 pile-up events and the PHS consisting solely of 1275 keV events.

However, since the signal overlap causes the pulse area and amplitude to sum, thereby extending the PHS towards higher values, we can take advantage of this effect to supress the contribution of these spectrum-distorting events by systematically lowering the upper threshold of the PHS start window.

Figure 3 shows a Gean4-simulated PHS [27, 28] using parameters derived from the experimental setup of Petschke et al. [16], similar to the one used in this study. As seen in the zoomed-in section, the pulse-height distribution related to 1275/511 pile-up events (orange dots) extends to much higher values (channels larger than 700) compared to the ideal PHS, where exclusively 1275 keV interactions are counted (green dots). Moreover, it can be observed that, starting from a certain pulse-height towards lower values in the PHS start window region, the relative contribution of the 1275/511 pile-up events remains nearly constant (below channel 550). Therefore, further shifting the upper PHS start window to lower values will not significantly improve their elimination rate. Instead, this would considerably reduce the efficiency, and using a 180° configuration would lose its main benefit.

Hence, finding an appropriate threshold value for the upper PHS start window, where efficiency is maintained while maximizing the elimination rate and, consequently, the spectrum quality, is crucial to this approach.

To validate this approach experimentally, the following sections explore:

- The impact of various PHS window settings on measured positron lifetimes with respect to the confidence in the relevant information extracted from the lifetime spectrum, by comparing it to the results obtained from a 90° configuration and to values of theoretical calculations known from the literature (Section 3.1).

- The influence of PHS start window selection on the IRF broadening (Section 3.2).

## 3. Results and Discussion

To experimentally validate our hypotheses regarding the reduction of corrupting 1275/511 pile-up events in a 180° configuration by adjusting the upper threshold of the PHS start window, we recorded a single lifetime spectrum of high-purity aluminum (5N5) using DDRS4PALS acquisition software (Version v1.18) [22].

DDRS4PALS software enables the recording of pulses as a binary stream, allowing for spectrum reprocessing under different software configurations, therefore permitting the generation of multiple spectra from a single measurement while only varying the upper PHS start window threshold. To ensure comparability between the spectra, each reprocessed measurement was stopped once the total count in the lifetime spectrum reached 4 million.

The experimental setup used was as described in [16]. For validation, a reference lifetime spectrum was obtained under identical experimental conditions in terms of sample material, source, scintillator and PMTs but in a 90° configuration where the detection of this studied event type is fully supressed.

The lifetime spectra were analysed using DQuickLTFit fitting software which uses the Levenberg-Marquardt algorithm for solving the non-linear least-square problem (Version v4.2) [29]. The employed fit model is the commonly used analytical solution of the convolution of a sum of N exponential distributions with a linear combination of gaussian distribution functions, as first published by Kirkegaard et al. [30]. The initial fit parameters were kept constant across all spectra, while all parameters were left free during the fitting process.

### *3.1. Influence on the spectra quality*

Table 1: Extracted lifetimes and intensities from measurements with 90° and 180° detector configurations using varying upper threshold of the PHS start window. $\tau_1$ and its corresponding intensity ($I_1$) represent the bulk lifetime of aluminum, while $\tau_2$, $\tau_3$, and their respective intensities ($I_2/I_3$) originate from the positron source.

|  | 90° setup | Upper threshold: 700 [channel] | Upper threshold: 440 [channel] |
|---|---|---|---|
| $\tau_1$ [ps] | 163.90 (0.80) | 158.86 (0.58) | 160.60 (0.58) |
| $I_1$ | 0.879 (0.006) | 0.870 (0.005) | 0.876 (0.005) |
| $\tau_2$ [ps] | 395.09 (4.87) | 379.25 (4.48) | 392.88 (4.80) |
| $I_2$ | 0.1154 (0.0005) | 0.1241 (0.0005) | 0.1184 (0.0005) |
| $\tau_3$ [ps] | 3344.50 (72.12) | 3081.72 (76.51) | 3170.63 (83.67) |
| $I_3$ | 0.0056 (0.0001) | 0.0056 (0.0001) | 0.0056 (0.0001) |

Figure 4 shows the extracted characteristic lifetime and their corresponding contribution for the bulk aluminum ($\tau_1$, $I_1$) alongside the two source components originating from the sodium sealed in Kapton foil ($\tau_2$, $I_2$ and $\tau_3$, $I_3$). Even though the third component is given in Figure 4 and Table 1, we won't refer to it as their contribution of less than one percent competes with the background and is therefore irrelevant for the validity of the study`s conclusions.

It can be noticed that the bulk and source components reveal a systematic dependence on the decreasing upper threshold of the PHS start window (channel: 700 → 440):

The bulk lifetime $\tau_1$ increases from 158.9±0.6 ps at a channel 700 to 160.6±0.6 ps at channel 440 while the source-related lifetime $\tau_2$ shows an increase from 379.3±4.5 ps to 392.9±4.8 ps. The intensity $I_1$, corresponding to the bulk component, increases from 0.870±0.005 to 0.876±0.005, whereas the intensity $I_2$, associated with the source component, exhibits a decrease from 0.1242±0.0005 to 0.1184±0.0005.

The lifetimes and intensities extracted from the spectrum recorded in the 90° configuration are considered more accurate, as the setup inherently suppresses DD events of the 511 keV gamma quanta due to its geometry of the detector pair. As a result, no influence from 1275/511 pile-up events is expected. This assumption is supported by comparison of the relevant PHS windows between the 90° and 180° configurations (see Figure 4), which exhibit the same difference as that observed between the ideal PHS and the realistic PHS in the Geant4 simulation shown in Figure 3. Furthermore, the characteristic lifetime attributed to bulk aluminium obtained with the 90° configuration (163.9±0.8 ps) shows a better agreement with theoretically calculated values, which range from 163 ps to 166 ps, as reported by various groups [31-33].

In Figure 4, the obtained lifetimes and intensities for the 90° configuration are shown, with coloured bars indicating the associated uncertainty bands. Additionally, for $\tau_1$, which represents the characteristic lifetime of bulk aluminium, the theoretically calculated values of various groups are indicated in pink.

As the upper threshold of the PHS start window is reduced, both the characteristic lifetimes and their corresponding intensities show a clear trend towards the values obtained with the 90° configuration. Notably, the source component's lifetime $\tau_2$ and the intensity $I_1$ of the bulk material converge fully within the fitting uncertainties. Although the remaining components do not reach the same absolute values, a consistent trend in that direction is evident. This behaviour results from the effective suppression of 1275/511 pile-up events; however, residual contributions from such coincidences remain within the reduced window range (see Figure 4), preventing the data from achieving the full accuracy of the 90° configuration.

Nonetheless, a critical balance must be struck between measurement time and spectrum quality. Reducing the upper threshold to channel 440 yields an efficiency loss of approximately 36% compared to the original, broader start window (see Figure 5 b), requiring proportionally longer

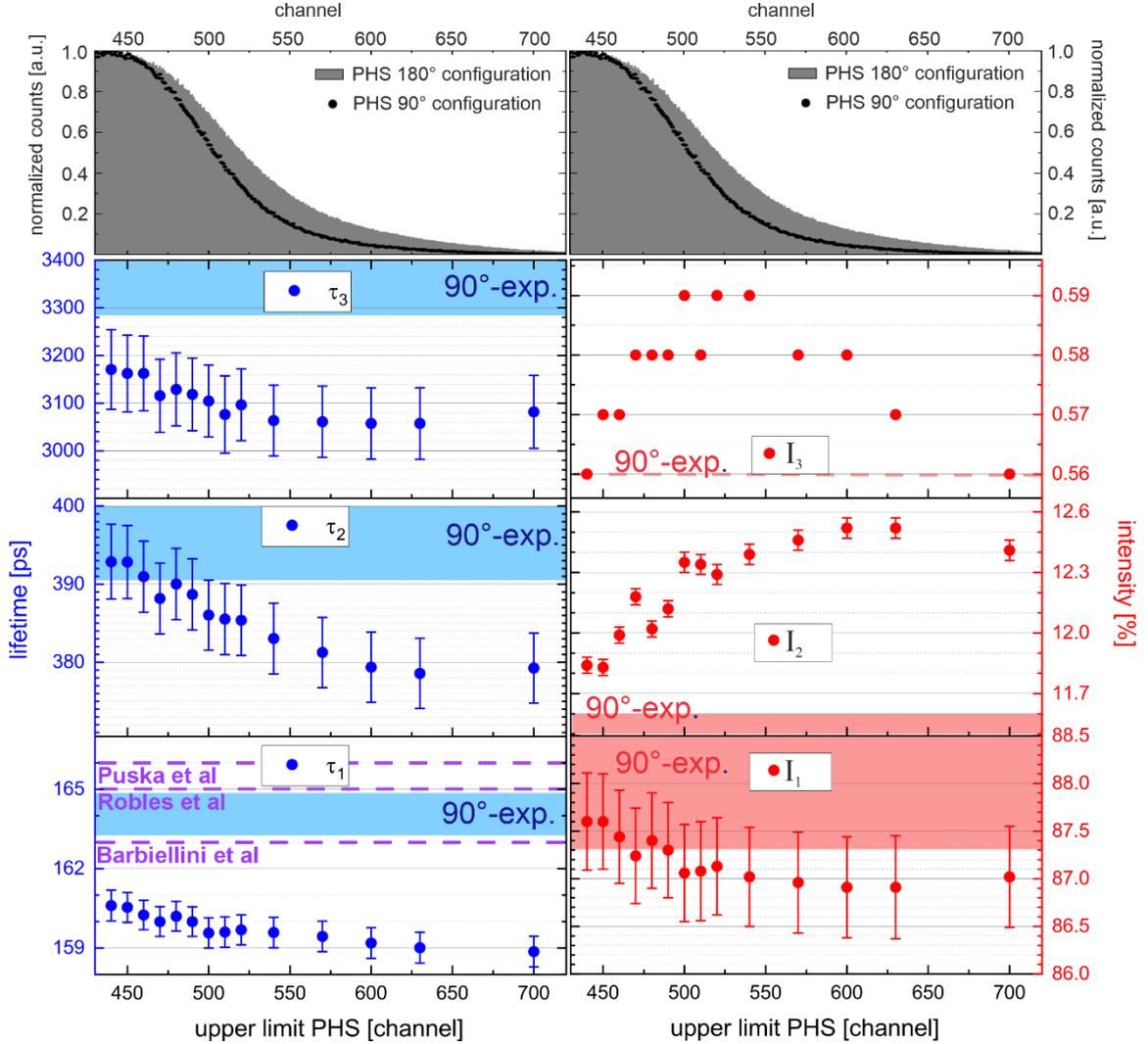

Figure 4 Influence of the reduction of the upper threshold of the PHS start window on the characteristic lifetimes and their corresponding contributions. The upper panel shows the corresponding section of the PHS that was systematically excluded. Extracted characteristic lifetimes (τ) of all three components are shown in blue as a function of the threshold value; corresponding intensities (I) are shown in red. The shaded band indicates the values from the 90° configuration including their uncertainties, labeled as *90°-exp*. In the bulk lifetime plot, three theoretically calculated reference values for aluminum are shown in pink. The values are taken from [31-33].

acquisition times. In contrast, the 90° configuration suffers from an intrinsic count rate reduction of about one order of magnitude (factor 10) compared to a standard 180° configuration. Despite the efficiency loss, the spectra obtained with the reduced upper threshold show significantly improved quality relative to the unfiltered configuration. These findings confirm that adjusting the upper threshold of the PHS start window can effectively suppresses corrupted 1275/511 pile-

– 8 –

up events, leading to a higher spectrum quality and thus a more reliable analysis without requiring hardware modifications.

## *3.2. Impact on the timing resolution*

In 1275/511 pile-up events, the START signal results from the overlap of two pulses (see Figure 2). Due to the altered pulse amplitude, time determination based on the CF principle is shifted to later times compared to a pure 1275 keV pulse. Consequently, a reduced time difference between the START and STOP signals is expected for such events.

Additionally, the transit time spread (TTS) of a PMT is known to follow an inverse square root dependence on the number of photoelectrons generated at the photocathode, TTS $\propto 1/\sqrt{n}$ [26]. Since 1275/511 pile-ups produce more scintillation light, and therefore more photoelectrons, the associated TTS, and thus the IRF, is narrower for these events.

However, at lower upper thresholds of the PHS start window, not only are 1275/511 pile-up events suppressed, but also true 1275 keV START signals with high light output. Consequently, this suppression contributes to a slight broadening of the IRF due to the reduction of events with intrinsically lower TTS.

As a result, reducing the upper threshold of the PHS start window leads to a broadening of the IRF's FWHM, due to the disproportionate removal of 1275/511 pile-ups and suppression of true high-light-yield 1275 keV events. Figure 5 a shows the FWHM of the IRF as a function of the upper threshold of the PHS start window. A slight increase in the FWHM by approximately 2 ps is observed as the start window threshold is lowered down to channel 440.

Nevertheless, this broadening remains negligible when compared to the substantial improvement in the quality of the extracted lifetime components (see Table 1). Since 1275/511 pile-ups introduce additional distortion in the lifetime spectrum, their suppression significantly enhances measurement reliability. The minor increase in the FWHM of the IRF is thus an acceptable trade-off for the improvement in spectral resolution.

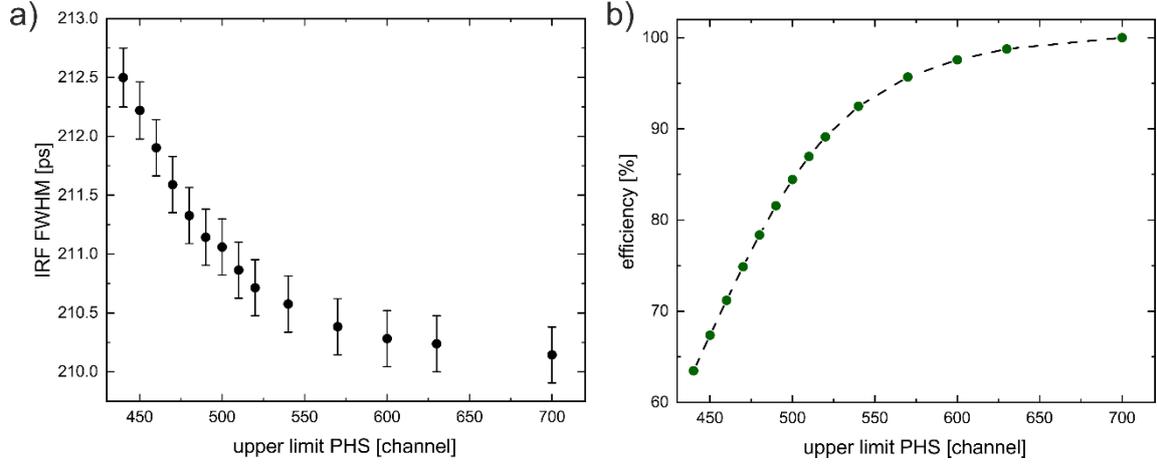

Figure 5: a) representation of the FWHM for the IRF of the measured lifetime spectrum as a function of varying upper limits of the start window in the PHS. b) shows the decrease in efficiency as a function of the lowering upper limit of the PHS start window threshold.

## 4. Conclusion

This study demonstrates that the quality of the positron lifetime spectrum and consequently the capability for a reliable and quantitative analysis of the investigated phenomena can be significantly enhanced by a systematic optimization of the upper threshold of the start window in the PHS, using a reference material of high-purity aluminium in a 180° detector configuration. This optimization strategy, which does not require any hardware modifications, effectively suppresses 1275/511 pile-up events.

By analysing the influence of different upper thresholds applied to the PHS start window in the measurement of high-purity (5N5) aluminium, the following key observations were made:

- The characteristic bulk lifetime of aluminium ($\tau_1$) exhibits a systematic increase with decreasing upper start window thresholds, approaching values obtained in a 90° detector configuration and aligning more closely with theoretically predicted bulk lifetimes reported in the literature.

- All other fitted parameters similarly tend to converge toward values observed in the 90° configuration. However, due to residual 1275/511 pile-up contributions within the reduced PHS start window, exact agreement within statistical uncertainties could not be achieved.

- The FWHM of the IRF exhibits only minor broadening (~2 ps) at decreasing upper start window thresholds, indicating a negligible impact on the overall timing resolution.

- The reduction of the upper start window threshold entails an efficiency loss of approximately 36%, necessitating an increased measurement duration by roughly one third compared to a configuration utilizing the full visible Compton edge in the PHS. Nonetheless, the optimized 180° configuration remains substantially more time efficient than the 90° configuration, while preserving high spectrum quality.

These findings indicate that a carefully optimized upper start window threshold of the PHS can partially compensate for the limitations of a 180° geometry, particularly in relation to distortions induced by 1275/511 pile-up events. As a result, extracted lifetime parameters can be brought significantly closer to those obtained in a 90° setup or theoretically expected ground truth values, where such coincidence distortions are inherently absent. Although the 90° configuration yields more reliable lifetime spectra, it requires considerably longer acquisition times, often in the order of one magnitude (factor 10) compared to the 180° configuration, rendering it impractical for routine measurements. The proposed approach thus offers a viable and efficient alternative, enhancing spectrum fidelity without compromising the intrinsic high detection efficiency associated with the 180° configuration.

## Acknowledgments

This research did not receive any specific grant from funding agencies in the public, commercial, or not-for-profit sectors.

## References

[1] D. Connors and R. West, *Positron annihilation and defects in metals.* Phys. Lett. A. **30** (1969) 24-25 doi: 10.1016/0375-9601(69)90018-8.

[2] P. Hautojärvi, A. Tamminen and P. Jauho, *Trapping of positrons by dislocations in aluminum.* Phys. Rev. Lett. **24** (1970) 459-461 doi: 10.1103/PhysRevLett.24.459.

[3] V. Crisp, D. Lock and R. West, *An investigation of positron trapping in indium.* J. Phys. F: Met. Phys. **4** (1974) 830 doi: 10.1088/0305-4608/4/6/011.

[4] A. Seeger, *The study of defects in crystals by positron annihilation.* Appl. Phys. **4** (1974) 183-199 doi: 10.1007/BF00884229.

[5] A. Seeger, *Investigation of point defects in equilibrium concentrations with particular reference to positron annihilation techniques.* J. Phys. F: Met. Phys. **3** (1973) 248 doi: 10.1088/0305-4608/3/2/003.

[6]     K. Saarinen, P. Hautojärvi and C. Corbel, *Positron annihilation spectroscopy of defects in semiconductors*, in *Semiconductors and Semimetals*. 1998, Elsevier. p. 209-285. doi: 10.1016/S0080-8784(08)63057-4.

[7]     R. Krause-Rehberg, et al., *Review of defect investigations by means of positron annihilation in II-VI compound semiconductors.* Applied Physics A: Materials Science & Processing. **66** (1998).

[8]     R. Krause-Rehberg and H.S. Leipner, *Positron annihilation in semiconductors: defect studies.* (1999).

[9]     G. Dlubek, et al., *Positron annihilation: a unique method for studying polymers*. in *Macromol. Symp.* 2004. Wiley Online Library. doi: 10.1002/masy.200450602.

[10]    S. Harms, et al., *Free volume of interphases in model nanocomposites studied by positron annihilation lifetime spectroscopy.* Macromolecules. **43** (2010) 10505-10511 doi: 10.1021/ma1022692.

[11]    K.O. Jensen and A.B. Walker, *Positron thermalization and non-thermal trapping in metals.* J. Phys. Condens. Matter. **2** (1990) 9757 doi: 10.1088/0953-8984/2/49/004.

[12]    M. Liu, B. Klobes and J. Banhart, *Positron lifetime study of the formation of vacancy clusters and dislocations in quenched Al, Al–Mg and Al–Si alloys.* Journal of Materials Science. **51** (2016) 7754-7767 doi: 10.1007/s10853-016-0057-7.

[13]    T. Nagoshi, et al., *Fatigue damage assessment of SUS316L using EBSD and PALS measurements.* Materials Characterization. **154** (2019) 61-66 doi: https://doi.org/10.1016/j.matchar.2019.05.038.

[14]    U. Ackermann, et al., *Time- and energy-resolution measurements of $BaF_2$, BC-418, LYSO and $CeBr_3$ scintillators.* Nucl. Instrum. Methods Phys. Res. A. **786** (2015) 5-11 doi: 10.1016/j.nima.2015.03.016.

[15]    F. Bečvář, et al., *The asset of ultra-fast digitizers for positron-lifetime spectroscopy.* Nucl. Instrum. Methods Phys. Res. A. **539** (2005) 372-385 doi: 10.1016/j.nima.2004.09.031.

[16]    D. Petschke, R. Helm and T.E.M. Staab, *Data on pure tin by Positron Annihilation Lifetime Spectroscopy (PALS) acquired with a semi-analog/digital setup using DDRS4PALS.* Data Brief. **22** (2019) 16-29 doi: 10.1016/j.dib.2018.11.121.

[17]    M. Fang, N. Bartholomew and A. Di Fulvio, *Positron annihilation lifetime spectroscopy using fast scintillators and digital electronics.* Nucl. Instrum. Methods Phys. Res. A. **943** (2019) 162507 doi: 10.1016/j.nima.2019.162507.

[18]    D. Boras, D. Petschke and T. Staab, *Geant4-based technical simulation study of plastic scintillators for Positron Annihilation Lifetime Spectroscopy (PALS).* Journal of Instrumentation. **20** (2025) T01009 doi: 10.1088/1748-0221/20/01/T01009.

[19]    K. Sedlak, A. Stoykov and R. Scheuermann, *A GEANT4 study on the time resolution of a fast plastic scintillator read out by a G-APD.* Nuclear Instruments and Methods in Physics Research Section A: Accelerators, Spectrometers, Detectors and Associated Equipment. **696** (2012) 40-45.

[20]    D. Bosnar, et al., *Triple coincidence PALS setup based on fast pulse digitizers.* Journal of Physics: Conference Series. **618** (2015) 012044 doi: 10.1088/1742-6596/618/1/012044.

[21]    L.Y. Dubov, et al., *Optimization of $BaF_2$ positron-lifetime spectrometer geometry based on the Geant4 simulations.* Nucl. Instrum. Methods Phys. Res. B. **334** (2014) 81-87 doi: 10.1016/j.nimb.2014.05.006.

[22]    D. Petschke and T.E.M. Staab, *Update (v1.3) to DLTPulseGenerator: A library for the simulation of lifetime spectra based on detector-output pulses.* SoftwareX. **9** (2019) 183-186 doi: 10.1016/j.softx.2019.02.003.

[23]    D. Boras, *Digital Positron-Annihilation-Lifetime-Spectroscopy: Geant4-Monte-Carlo-Simulation and Pulse-Generator-Coupling for a Realistic Digital Twin*. 2024, University Wuerzburg.

[24] F. Bečvář, *Methodology of positron lifetime spectroscopy: Present status and perspectives.* Nucl. Instrum. Methods Phys. Res. B. **261** (2007) 871-874 doi: 10.1016/j.nimb.2007.03.042.

[25] Q.H. Zhao, et al., *A multi-parameter discrimination digital positron annihilation lifetime spectrometer using a fast digital oscilloscope.* Nucl. Instrum. Methods Phys. Res. A. **1023** (2022) 165974 doi: 10.1016/j.nima.2021.165974.

[26] D. Petschke and T.E.M. Staab, *DDRS4PALS: A software for the acquisition and simulation of lifetime spectra using the DRS4 evaluation board.* SoftwareX. **10** (2019) 100261 doi: 10.1016/j.softx.2019.100261.

[27] S. Agostinelli, et al., *GEANT4—a simulation toolkit.* Nucl. Instrum. Methods Phys. Res. A. **506** (2003) 250-303 doi: 10.1016/S0168-9002(03)01368-8.

[28] J. Allison, et al., *Geant4 developments and applications.* IEEE Transactions on nuclear science. **53** (2006) 270-278 doi: 10.1109/TNS.2006.869826.

[29] D. Petschke, *DQuickLTFit - A least-square fitting tool for the analysis of positron lifetime spectra using the Levenberg-Marquardt algorithm.* 2021 (available: https://github.com/dpscience/DQuickLTFit?tab=readme-ov-file). doi: 10.5281/zenodo.1168285.

[30] P. Kirkegaard and M. Eldrup, *POSITRONFIT: A VERSATILE PROGRAM FOR ANALYSING POSITRON LIFETIME SPECTRA*. 1972, Danish Atomic Energy Commission Research Establishment, Risoe.

[31] M. Puska, *Ab-initio calculation of positron annihilation rates in solids.* Journal of Physics: Condensed Matter. **3** (1991) 3455 doi: 10.1088/0953-8984/3/20/007.

[32] B. Barbiellini, P. Genoud and T. Jarlborg, *Calculation of positron lifetimes in bulk materials.* J. Phys. Condens. Matter. **3** (1991) 7631 doi: 10.1088/0953-8984/3/39/009.

[33] J.C. Robles, E. Ogando and F. Plazaola, *Positron lifetime calculation for the elements of the periodic table.* J. Phys. Condens. Matter. **19** (2007) 176222 doi: 10.1088/0953-8984/19/17/176222.